\documentclass[fleqn,usenatbib]{mnras}

\usepackage{newtxtext,newtxmath}

\usepackage[T1]{fontenc}
\usepackage{ae,aecompl}

\usepackage{graphicx}	
\usepackage{amsmath}	
\usepackage{amssymb}	

\usepackage{pifont}

\usepackage{color, colortbl}
\definecolor{Gray}{gray}{0.9}

\title[Metre-scale Earth impactors]{Observation of metre-scale impactors by the Desert Fireball Network}

\author[H. A. R. Devillepoix et al.]{
H. A. R. Devillepoix,$^{1}$\thanks{E-mail: h.devillepoix@postgrad.curtin.edu.au}
P. A. Bland,$^{1}$
E. K. Sansom,$^{1}$
M. C. Towner,$^{1}$
M. Cup{\'a}k,$^{1}$
\newauthor
R. M. Howie,$^{1}$
B. A. D. Hartig,$^{1}$
T. Jansen-Sturgeon,$^{1}$
M. A. Cox$^{1}$
\\
$^{1}$School of Earth and Planetary Sciences, Curtin University, GPO Box U1987, Perth WA 6845, Australia
}

\date{Accepted XXX. Received YYY; in original form ZZZ}

\pubyear{2018}

\begin{document}
\label{firstpage}
\pagerange{\pageref{firstpage}--\pageref{lastpage}}
\maketitle

\begin{abstract}
	The Earth is impacted by 35-40 metre-scale objects every year.
	These meteoroids are the low mass end of impactors that can do damage on the ground. Despite this they are very poorly surveyed and characterised, too infrequent for ground based fireball observation efforts, and too small to be efficiently detected by NEO telescopic surveys whilst still in interplanetary space.
	We want to evaluate the suitability of different instruments for characterising metre-scale impactors and where they come from.
    We use data collected over the first 3 years of operation of the continent-scale Desert Fireball Network, and compare results with other published results as well as orbital sensors.
    We find that although the orbital sensors have the advantage of using the entire planet as collecting area, there are several serious problems with the accuracy of the data, notably the reported velocity vector, which is key to getting an accurate pre-impact orbit and calculating meteorite fall positions. We also outline dynamic range issues that fireball networks face when observing large meteoroid entries.
\end{abstract}


\section{Introduction}
	The Earth is impacted by 35-40 metre-scale objects every year \citep{2002Natur.420..294B,2006M&PS...41..607B}.
	These large meteoroids are at the low mass end of potentially damage-causing impacting asteroids like Chelyabinsk \citep{2013Natur.503..238B}.
	The study of the atmospheric behaviour, physical nature, numbers, and dynamical origin of these objects is therefore important in order to assess the hazard they pose, and prepare an appropriate response should an asteroid be detected and determined to be on a collision course with earth.

	\subsection{How frequently do these impacts happen?}
	One of the ways the size frequency distribution (SFD) of metre-scale has been surveyed is by using the so-called US Government (USG) sensors\footnote{\url{https://cneos.jpl.nasa.gov/fireballs/} accessed November 22, 2017}, which are able to detect flashes all around the world, day and night, measure flash energy, and sometimes derive velocities and airburst heights.
	As outlined by \citet{2013Natur.503..238B}, there might be subtleties in the SFD, namely a larger number of 10-50\,m objects. Indeed the 1-100\,m size range is largely unobserved, with objects too small for telescopes and too infrequent for impact monitoring systems to get representative surveys.
	So far, there have been 3 cases of asteroids detected before atmospheric impact. These are asteroids 2008 TC3 \citep{2009Natur.458..485J, 2017Icar..294..218F}, 2014\,AA \citep{2016Icar..274..327F}, and 2018\,LA, all discovered by the Catalina Sky Survey only hours before impact. As large deep surveyors like LSST \citep{2008SerAJ.176....1I} come online these types of detections are going to become more common, and predicting the consequences of these impacts is going to be desirable. While the impact location of 2008\,TC3 was well constrained to sub kilometre precision thanks to a very large number ($\simeq$900) of astrometric measurements, the prediction for 2014\,AA was much more uncertain and covered a large area of the Atlantic ocean, as only a total of 7 astrometric positions were available.
	The impact location of 2018\,LA was very uncertain, until 2 extra observation by the Asteroid Terrestrial-impact Last Alert System (ATLAS) increased the observation arc length from 1.3\,hours to 3.7 hours, which narrowed down the impact location to South Africa.	
	The number of astrometric observations and the length of the observation arc are therefore a critical factors to precisely determining the impact point. Well coordinated, large follow-up networks of telescopes can provide large numbers of such observations and  will aid in future impact predictions \citep{2016DPS....4840506L}.

	\subsection{How dangerous are these impacts?}
	The damage from an impact depends not only on dynamical parameters, but also on: size, rock type, structure, strength ($s$) and density ($\rho$).
	To illustrate this, we can use the equations of \citet{2005M&PS...40..817C} to simulate the outcome of the impact of a 2\,m object, with an entry angle of 18\degr, a velocity of 19\,$\mbox{km s}^{-1}$ at the top of the atmosphere (same entry angle and velocity as Chelyabinsk), and various bulk strengths and densities corresponding to different classes of objects (from \citet{1993Natur.361...40C}):
	\begin{itemize}
		\item a weak cometary body ($s=10^5$\,Pa, $\rho=1000\,\mbox{kg m}^{-3}$) will break up at a high altitude (60\,km), causing no significant direct damage because the predicted 0.18\,kT TNT of energy released cannot be transferred efficiently to the ground due to the thin atmosphere (1\,kT\,TNT = $4.184\times10^{12}$\,J).
		\item a chondritic body ($s=10^7$\,Pa, $\rho=3500\,\mbox{kg m}^{-3}$) is likely going to airburst at relatively low altitudes (the model predicts an airburst at 27\,km), releasing around 0.44\,kT TNT of energy that can be propagated more efficiently by the denser atmosphere.
		\item an iron ($s=10^8$\,Pa, $\rho=7900\,\mbox{kg m}^{-3}$) monolith will reach the surface at hypersonic velocity (3.8\,$\mbox{km s}^{-1}$), causing important but very localised damage, as it only yields $10^{-1}$\,kT TNT.
	\end{itemize}
	This is a simplistic example, but it shows how much the response to an imminent asteroid impact depends on both physical and dynamical characteristics of the impactor.
	
	Several observation techniques can be levied while the asteroid is still in interplanetary space:
	\begin{itemize}
		\item Multi-band photometry in Vis-NIR: size and rotation period, and lower constraint on cohesive strength as a consequence.
		\item Spectroscopy: likely composition.
		\item Astrometric observations: pre-encounter orbit, and predictions about the impact geometry, velocity, and location.
		\item Radar observations: size, shape, rotation period, presence of satellites.
	\end{itemize}

	While the size and impacting velocity are well constrained factors using astrometric observations, determining the rock type and structure from remote sensing instruments is more challenging.
	
	To some extent spectroscopy can provide insights on the mineralogy of the impactor, but this technique requires a good knowledge of how asteroid spectral types match meteorite types.
	
	Another approach is the work of \citet{2014ApJ...786..148M,2014ApJ...789L..22M} on small (metre-scale) asteroids for which spectroscopic work is generally impractical. They used a thermophysical model combined with an orbital model that takes non-gravitational forces into accounts. This model derives physical parameters (likely surface composition, size) by combining both astrometric observations and Near-Infrared photometry. 
	
	In order to be reliable on large scales, these techniques have to be qualified with direct sample analysis.
	This active area of research can be tackled in two ways: either direct sample return missions (like Stardust, Hayabusa, Hayabusa 2, OSIRIS-REx), or from a large number of meteorite recoveries with  associated orbits that can link to asteroid families: the aim of ground-based efforts like the Desert Fireball Network.
	
	The Desert Fireball Network (DFN) is a fireball camera network currently operating in the the Australian outback, designed for the detection and recovery of meteorite falls with associated orbits. Currently 52 observatories are deployed.
	On January 2, 2015, a particularly bright fireball was observed over South Australia, large enough to be simultaneously detected by the US government (USG) sensors, and by the DFN, which had just started science operation 2 months before. Another similarly bright event, also observed by both the DFN and the USG sensors, happened on June 30, 2017 over South Australia.
	
	Over the 3 million km$^2$ that the DFN covers in Australia, the observation of a metre-scale impactor is only expected to happen once every 4-5 years \citep{2002Natur.420..294B}, and once every 8-10 years during night time when most dedicated fireball networks operate (without considering clear sky conditions). The observation of two such events during the first 3 years of operation of the DFN, although outside the nominal collecting area, is somewhat lucky with respect to the size frequency distribution numbers of \citet{2002Natur.420..294B}.
	These two superbolides are described here and add to the small list of metre-scale impactors that have precisely determined trajectories:
	\begin{itemize}
		\item 13 events compiled and discussed by \citet{2016Icar..266...96B}.
		\item the "Romanian" bolide \citep{2017P&SS..143..147B}.
		\item the Dishchii'bikoh meteorite, for which initial trajectory details have been reported by \citet{2018arXiv180105072P}.
		\item the meteorite fall near Crawford Bay in British Columbia (Canada), for which initial trajectory details have been reported by \citet{2018LPI....49.3006H}. 
	\end{itemize}

	\subsection{Where do they come from?}
	The current state of the art for source region model for Near-Earth Objects (NEO) is detailed by \citet{2018Icar..312..181G}. They report a significant size dependence of NEO origins, which had not been investigated by earlier similar works \citep{2002Icar..156..399B,2004Icar..170..259B,2012Icar..217..355G}.
	Their work covers the absolute magnitude range $17<H<25$ (corresponds to diameter $1200>D>30$\,m with an S-type albedo of 0.2), providing little insight on the the metre-size region ($H=32$).
		
	Several outstanding issues show that it is not possible to simply interpolate the characteristics of the population of typical macroscopic meteorite dropper meteoroids (decimetre-scale) and the kilometre-scale  well surveyed by telescopes.
	For instance, LL chondrites make up 8\% of meteorite falls, but it is generally thought that $1/3^{rd}$ of observable near-earth small body space is made up of LL compatible asteroids \citep{2008Natur.454..858V}.
	\citet{2016Natur.530..303G} shows that an unmodelled destructive effect prevents small bodies from stably populating the low perihelion region, further outlying the need to consider body size in the dynamical models.
	
	\citet{2016Icar..266...96B} are the first to perform a source region analysis on metre-class NEO bodies, using the \citet{2002Icar..156..399B} model on USG events.
	Considering the small number statistics they get intermediate source regions proportion that are comparable to previous works on kilometre-size NEO population \citep{2002Icar..156..399B,2004Icar..170..259B,2012Icar..217..355G}.
	However they also argue for a Halley-type comet (HTC) source region, comparable in importance to the Jupiter-family comets (JFC) source.
	This source has not been identified previously in NEO works, because of a near-complete lack of such objects in asteroid databases.
	Their argument is based on three fireball events in the USG dataset that have a Tisserand parameter with Jupiter, $T_J<2$: identified as \textit{20150102-133919}, \textit{20150107-010559}, and \textit{20150311-061859}, not associated to a meteor shower.
	Because the first two of these events have independently estimated trajectories, 
	an issue that we are interested in is determining if this surprising outcome could be the results of limitations of USG data.
	
	This work aims to compile independent information not just for these cases, but for several other metre-scale bodies, to determine the reliability of USG data in general, for population study, orbit determination, as well as undertaking meteorite searches based on these data.
	We also evaluate the suitability of hardware currently deployed by fireball networks to observe these particularly bright events.

	\section{Data and methods}
	
	\subsection{Desert Fireball Network}\label{sec:methods_dfn}
	The Desert Fireball Network (DFN) is the world's biggest fireball observation facility (3 million km$^2$ coverage), set up in a desert environment where meteorites are more likely to be successfully recovered.
	The DFN is built to overcome the challenges of operating a distributed network of high technology devices in a harsh remote environment.
	The observatories operate completely autonomously for up to two years before maintenance is required: swapping the hard drives and replacing the mechanical shutter in the off-the-shelf camera.
	The systems can operate with network connectivity for event notifications, or completely offline. Due to their low power usage, simple solar photo-voltaic systems ($\approx$160-240 W of solar panels) with 12 V deep-cycle lead acid battery storage are used to power most of the observatories across the network.
	
	The main imaging system consists of a high-resolution digital camera and a fisheye all-sky lens, taking long exposures with shutter breaks embedded by the GNSS synchronised operation of a liquid crystal shutter.
	This mode of imaging has historically been the most successful method for determining positions of fallen meteorites from fireball observation, as shown in the compilation of \citet{2015aste.book..257B}.
	The DFN has recovered 3 meteorites in the first 3 years of operation \citep{2018arXiv180302557D}.
	The automated observatories are more completely described by \citet{2017ExA...tmp...19H}, and the encoding method used to record absolute and relative timing (to derive velocity information) is detailed by \citet{2017M&PS...52.1669H}.
	
	In June 2017, the DFN initiated a firmware upgrade across the network to change the time encoding technique on the observatories' microcontroller. These were deployed to all online cameras remotely.
	The main new feature of this update was a new mode of operation for the liquid crystal shutter, different from the one described by \citet{2017M&PS...52.1669H}.
	This new mode retained the absolute timing encoding through the use of a de Bruijn sequence, but made the pulses much shorter and equal in duration, replacing the 60\,ms long dash with two 10\,ms pulses and the short 20\,ms dash into a single 10\,ms pulse, in order to reduce saturation issues on bright fireballs, and make automated centroid determination easier. In Tab. \ref{table:stations_Kalabity} and \ref{table:stations_BairdBay} we refer to this new method as pulse-frequency ($PF$), as opposed to the pulse-width ($PW$) method of \citet{2017M&PS...52.1669H}.
	
	Standard data reductions methods are detailed by \citet{2018arXiv180302557D}.
	The DFN is optimised to observe macroscopic meteorite dropping events at the low mass end. The observatories are sensitive to apparent magnitude 0, in order observe a small ($\sim5$\,cm) meteoroid high-enough before significant atmospheric deceleration happens, to derive a precise orbit. But they can also astrometrically observe the brightest phases of ablation of a half-metre size rock (magnitude 15), albeit with saturating the sensor.
	
	Thanks to the large number of stars imaged by the long exposure, the cameras typically achieve their nominal arcminute astrometric precision down to 5\degr elevations above the horizon \citep{2018arXiv180302557D}.
	Typical kilogram scale meteorites usually ablate down to $\sim 20$\,km height, therefore the network is spaced in order to have 3 camera observation down to this height, which roughly corresponds to a $200$\,km slant range.
	Outside of these ideal observation conditions, fireballs are accurately imaged in the high altitude phase of the flight (useful for orbital calculations), but getting precise meteorite fall positions becomes more difficult due to decreased astrometric precision.
	
	Fireball trajectories are calculated using a modified version of the least-square method of \citet{1990BAICz..41..391B}, and fireball dynamics are analysed using the methods of \citet{2015M&PS...50.1423S} and \citet{2017ASSP...46..153G}.
	Pre-encounter orbits are determined using numerical integration, as described by \citet{2018arXiv180805768J}.
	
	The DFN observatories were designed with a low-resolution video system in parallel of the high-resolution still imager, initially as absolute timing device, but later kept on some systems for future daytime observations.
	These data are too low-resolution to provide useful astrometric data, although they can be helpful in getting high temporal resolution photometric data.
	However the sensor gets saturated when the fireball gets brighter than $m_V=-5$, and the auto-gain on the cameras can only attenuate the signal by a factor of about 4 stellar magnitudes. Large fireballs still saturate the sensor, however \citet{2018arXiv180302557D} have successfully used the sum of all pixels in each field as a proxy for all sky brightness. This method is particularly successful at detecting large fragmentation events.
	The effect of auto-gain are corrected by performing traditional photometry on a non saturated bright star, planet, or fixed light in the field of view.
	Unfortunately because of the lossy compression of the record and the sensor saturation, it is not possible to get a satisfying absolutely calibrated photometry from the video, and therefore the resulting light curve is only used qualitatively.

	\begin{figure*}
		\centering
		\includegraphics[width=1\linewidth]{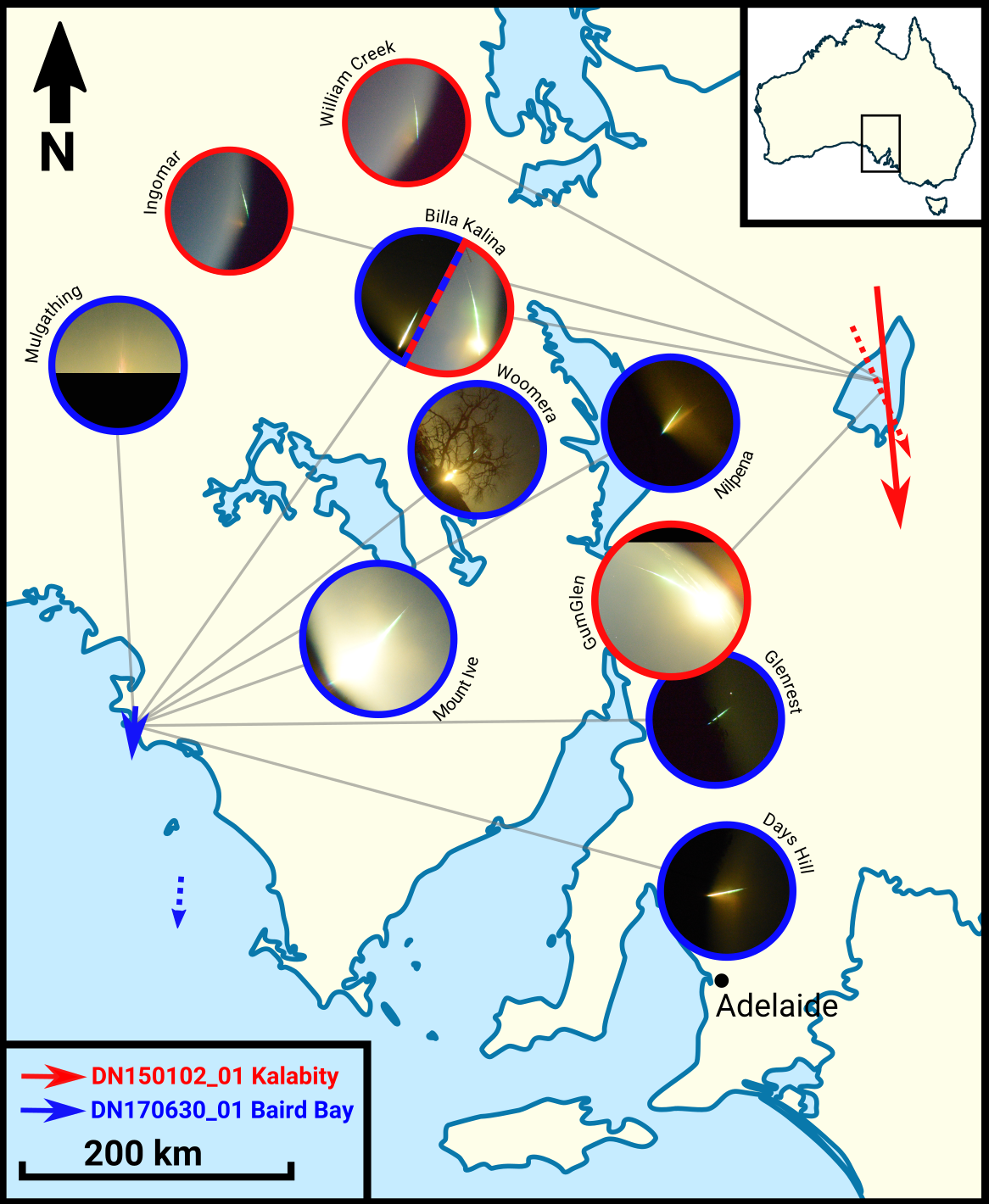}
		\caption{Cropped all-sky images of the fireballs from the DFN observatories. Images are of the same pixel scale with the centre of each image positioned at the observatory location on the map.
			For the Kalabity fireball (red arrow, East), light from the main explosion is particlaurly scattered in the Gum Glen image because of clouds.
			For the Baird Bay event (blue arrow, West), the Mulgathing image is cropped because the sensor is not large enough to accommodate the full image circle on its short side. The fireball on the Woomera picture was partially masked by a tree. The O'Malley station only recorded video and is missing in this map, detais are given in Tab. \ref{table:stations_BairdBay}.
			The dashed arrows show the USG sensors trajectory solutions for both events (vectors are generated by backtracking the state vector at the time of peak brightness to $t-5\,$seconds).}

		\label{fig:images_map}
	\end{figure*}

	\subsection{US government sensors}\label{sec:usg_methods}
	
	Large fireballs detected by the so-called "US Government (USG) sensors" are reported on the JPL website\footnote{\url{https://cneos.jpl.nasa.gov/fireballs/} accessed November 22, 2017}. These sensors are apparently able to detect flashes all around the world, day and night, measure flash energy, and sometimes derive velocities and airburst heights. These data were used for size-frequency studies of metre-scale objects by \citet{2002Natur.420..294B}, and later to derive orbital and physical properties of this population \citep{2016Icar..266...96B}.
	
	In Tab. \ref{table:usg_data} we give the data for a the subset of events for which all the parameters are reported (time, energy, location, velocity), and for which independent observations have been published (references in Tab \ref{table:usgscomp}).
	The USG sensors data do not come with uncertainties, therefore we assume the last significant figure represents the precision of the measurement.
	
	We calculate the radiant and pre-entry orbits for these meteoroids, based on USG data, using the numerical method of \citet{2018arXiv180805768J}.
	The various numbers reported in USG data relate to the instant of peak brightness, typically quite deep into the atmosphere.
	Since we are dealing with metre-scale bodies, we ignore deceleration due to the atmosphere and use a purely gravitational model from that point for calculating the orbit.
	
	The online table converts the total radiated energy measured into an equivalent impact energy using an empirical relation determined by \citet{2002Natur.420..294B}.
	This total energy estimate, combined with the impacting speed, can be used to derive a photometric mass using the classical kinetic energy relation ($E=\tfrac{1}{2}mv^2$), and a rough size assuming a density.
	
	Only metre-sized and larger objects are reported by the USG\footnote{Johnson L. (2017) - SBAG meeting: \url{https://www.lpi.usra.edu/sbag/meetings/jan2017/presentations/Johnson.pdf} and remarks at 32m and answer to questions at 56m in online talk: \url{https://ac.arc.nasa.gov/p98hreesxa9/}}, which roughly corresponds to a $0.1$\,kT TNT impact energy.

	\begin{table*}[h!]
		\caption{USG sensors fireball events that have reported velocities, and have been observed independently. Left of the table is from \url{https://cneos.jpl.nasa.gov/fireballs/} (accessed November 22, 2017). Apparent radiants and orbits have been calculated (angles are equinox J2000). The number of decimals is not representative of uncertainty. A similar work has been done by \citet{2016Icar..266...96B} on the 6 older events in the list, in agreement. Corresponding ground based observation details of the two highlighted events are presented in this work, references for the other events are in Tab. \ref{table:usgscomp}.}
		\label{table:usg_data}      
		
		\scalebox{0.8}{
			\begin{tabular}{cccccccccc|ccccccc}
				\hline \hline
				peak brightness & H & V & $V_{x}$ & $V_{y}$ & $V_{z}$ & radiated E. & impact E. & lat & long & $\alpha_{inf}$ & $\delta_{inf}$ & $a$ & $e$ & $i$ & $\omega$ & $\Omega$ \\
				date/time UT & km & $\mbox{km s}^{-1}$  & & $\mbox{km s}^{-1}$ & ECEF & J & kT TNT & \degr N+ & \degr E+ & \degr & \degr & AU &  & \degr & \degr & \degr \\ \hline
				2018-06-02 16:44:12 & 28.7 & 16.9 & 0.9 & -16.4 & 3.9 & 3.8e+11 & 0.98 & -21.2 & 23.3  & 235.0 & -13.3 & 1.33 & 0.42 & 4.4 & 258.3 & 71.88 \\
				2017-09-05 05:11:27 & 36.0 & 14.7 & 12.7 & -6.1 & -4.2 & 3.8e+10 & 0.13 & 49.3 & 243.1 & 216.5 & 16.7 & 2.06 & 0.54 & 3.4 & 147.2 & 162.69 \\
				\rowcolor{Gray} 2017-06-30 14:26:45 & 20.0 & 15.2 & 10.9 & -9.7 & 4.2 & 9.4e+10 & 0.29 & -34.3 & 134.5 & 273.6 & -16.1 & 1.24 & 0.35 & 3.6 & 259.7 & 98.80 \\
				2015-01-07 01:05:59 & 45.5 & 35.7 & -35.4 & 1.8 & -4.4 & 1.4e+11 & 0.4 & 45.7 & 26.9 & 119.7 & 7.1 & 4.78 & 0.93 & 20.7 & 111.9 & 106.19 \\
				\rowcolor{Gray} 2015-01-02 13:39:19 & 38.1 & 18.1 & 4.5 & -14.4 & -10.0 & 2.0e+10 & 0.073 & -31.1 & 140.0 & 53.8 & 33.5 & 9.33 & 0.90 & 8.0 & 207.4 & 281.61 \\
				2013-02-15 03:20:33 & 23.3 & 18.6 & 12.8 & -13.3 & -2.4 & 3.8e+14 & 440.0 & 54.8 & 61.1 & 329.2 & 7.3 & 1.71 & 0.56 & 4.1 & 109.7 & 326.46 \\
				2010-02-28 22:24:50 & 37.0 & 15.1 & -11.7 & 2.7 & -9.1 & 1.5e+11 & 0.44 & 48.7 & 21.0 & 121.7 & 37.2 & 2.70 & 0.65 & 3.2 & 204.0 & 340.14 \\
				2008-11-21 00:26:44 & 28.2 & 12.9 & 3.9 & -4.1 & -11.6 & 1.4e+11 & 0.41 & 53.1 & 250.1 & 200.6 & 64.0 & 0.79 & 0.26 & 10.4 & 3.1 & 239.04 \\
				2008-10-07 02:45:45 & 38.9 & 13.3 & -9.0 & 9.0 & 3.8 & 4.0e+11 & 1.0 & 20.9 & 31.4 & 12.4 & -16.7 & 1.65 & 0.43 & 3.9 & 36.2 & 14.16 \\
				\hline
		\end{tabular}}
	\end{table*}

	\newpage

	\section{Results}
	
	In this section we analyse in detail the atmospheric entry of 2 large meteoroids as observed by the DFN, these were also observed by the USG sensors (highlighted rows in Tab. \ref{table:usg_data}).
	
	\subsection{DN150102\_01 - Kalabity}\label{sec:kalabity_results}
	
\begin{figure*}
		\centering
		\includegraphics[width=\linewidth]{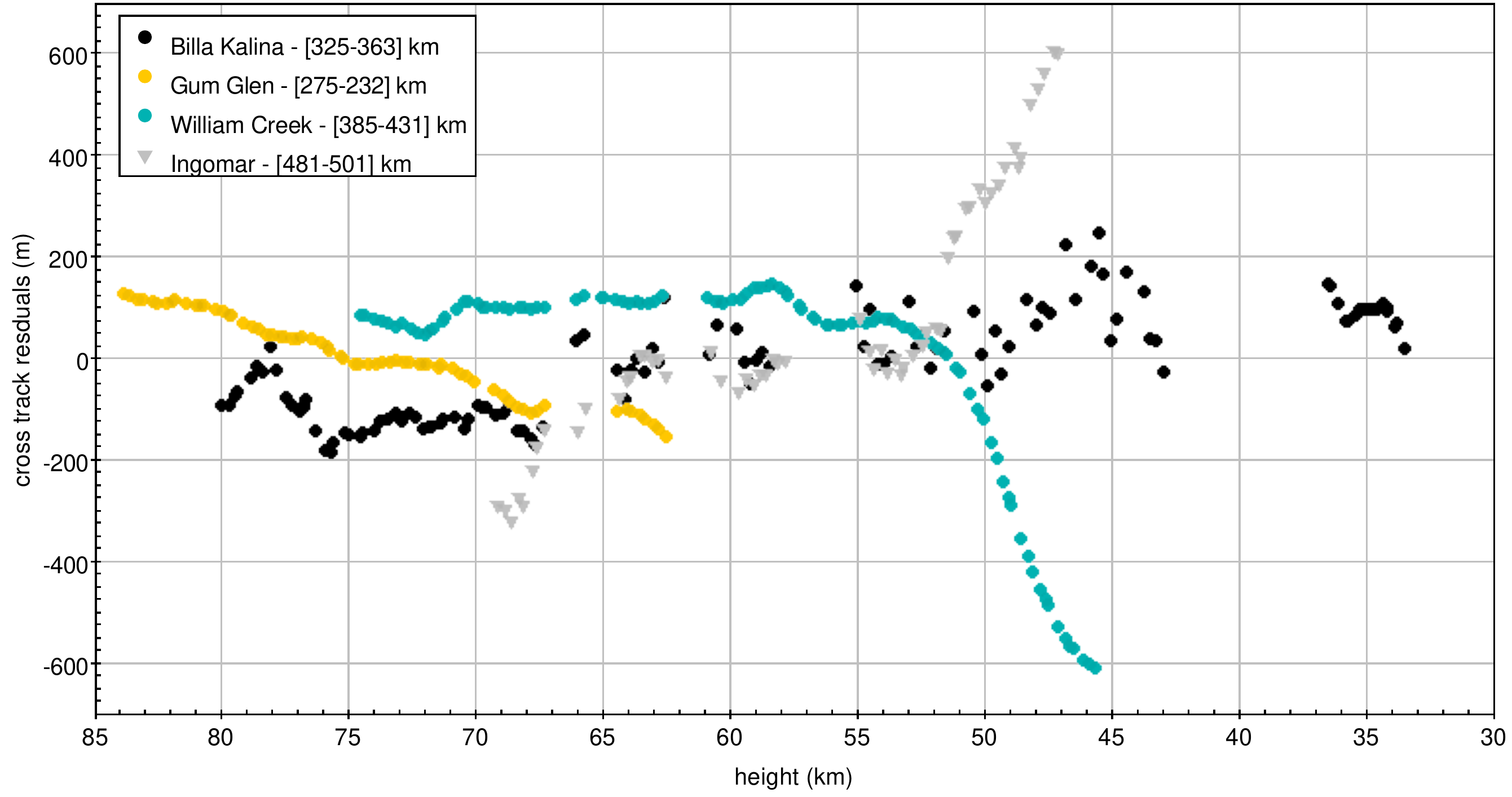}
		\caption{Cross-track residuals of the straight line least squares fit to the trajectory from each view point. These distances correspond to astrometric residuals projected on a perpendicular plane to the line of sight, positive when the line of sight falls above the trajectory solution. The distances in the legend correspond to the observation range [highest point - lowest point]. The Ingomar and William Creek observation residuals start diverging after 52\,km altitude, this corresponds to observation elevation angles of about 4\degr and 5\degr, respectively.}
		\label{fig:Kalabity_residuals}
	\end{figure*}

	On January 2, 2015 a bright bolide lit up the skies over lake Frome in South Australia (Fig. \ref{fig:images_map}), starting at 2015-01-02T13:39:11.086 UTC (9 minutes after midnight ACDT) for 10.54 seconds.
	In early 2015 the DFN had just finished its initial expansion phase in South Australia with 16 cameras, unfortunately the bolide happened outside the standard network covering area at that time. Therefore a combination of cameras mostly over 300\,km from the event had to be used to determine the trajectory (Tab. \ref{table:stations_Kalabity}). The best convergence angle is 22\degr (between Gum Glen and William Creek).
	The convergence angle between the Billa Kalina and Ingomar stations is less then 1\degr, therefore the latter distant viewpoint does not help much in constraining the trajectory.
	The trajectory follows a relatively shallow slope of 20\degr to the horizon, visible on the images from 83.3\,km altitude.
	Astrometric uncertainties vary between 1.5-3\arcmin (equates to 130-260\,m once projected at 300\,km). These are obtained by compounding astrometric calibration uncertainties (typically 1\arcmin) and fireball picking uncertainties (usually 0.5-1 pixel, depending on optics quality and fireball brightness).
	Most of the residuals to the straight line fit (Fig. \ref{fig:Kalabity_residuals}) are then in agreement with astrometric uncertainties.
	As expected from an unconstrained astrometric solution under 5\degr elevation, the observation residuals to the straight line fit start diverging for observations below this elevation, this is visible on  around the 52\,km altitude mark on the Ingomar and William Creek viewpoints.

	\begin{table*}
		\caption{Locations and nature of instrumental records DN150102\_01. P: Photographic record (long-exposure high resolution image), V: compressed PAL video (25 frames per second). $PW$ designates the de Bruijn encoding method, as described in Sec. \ref{sec:methods_dfn}. Ranges are from the fireball at 70\,km altitude. Photographic imaging system was out of order for Nilpena.}              
		\label{table:stations_Kalabity}      
		\centering                                      
		\begin{tabular}{c c c c c c}          
			\hline\hline                        
			Observatory & Instruments & Latitude & Longitude & Altitude (m) & Range (km) \\
			\hline      
			Gum Glen - DFNSMALL25 & P$_{PW}$, V &  32.20554 S & 138.24121 E & 242  & 246  \\ 
			Billa Kalina - DFNSMALL26 & P$_{PW}$ &  30.23769 S & 136.51565 E & 114  & 328  \\ 
			William Creek - DFNSMALL30 & P$_{PW}$ &  28.91566 S & 136.33495 E & 79  & 392  \\ 
			Ingomar - DFNSMALL27 & P$_{PW}$ &  29.58556 S & 135.03865 E & 197  & 480  \\ 
			Nilpena - DFNSMALL42  & V &  31.02331 S & 138.23256 E & 112  & 175  \\  
			\hline                                             
		\end{tabular}
	\end{table*}
	
	\begin{table*}
		\centering
		\begin{tabular}{lcccccc}
			\hline
			Event & Time &  Speed &  Height &  Longitude &  Latitude &  Dynamic pressure \\
			& s & $\mbox{m s}^{-1}$ & m & \degr E & \degr N & MPa \\
			\hline
			Beginning & 0.0 & 15406$\pm$79 & 83317 & 139.73897 & -30.25421  & \\
			A & 3.90 & 15351 & 62586 & 139.85081 & -30.74874 & 0.05 \\
			B & 4.50 & 15320 & 59453 & 139.86679 & -30.82416 & 0.08 \\
			C & 7.61 & 14487 & 43432 & 139.95010 & -31.21547 & 0.52 \\
			D & 7.83 & 14272 & 42571 & 139.95466 & -31.23679 & 0.57 \\
			E - max & 8.55 & 13463 & 40286 & 139.96683 & -31.29360 & 0.69 \\
			F & 8.95 & 13014 & 39017 & 139.97359 & -31.32517 & 0.77 \\
			G & 9.26 & 12665 & 38033 & 139.97883 & -31.34963 & 0.83 \\
			End & 10.54 & 8433 & 33420& 140.00311 & -31.46438 & \\
			\hline
		\end{tabular}
		\caption{Summary table of bright flight events for DN150102\_01 Kalabity. Fragmentation event letters are defined on the light curve (Fig. \protect\ref{fig:kalabity_radiometric_light_curve}). Times are relative to 2015-01-02T13:39:11.086\,UTC. Positions and speeds at the peaks are interpolated from astrometric data.}
		\label{tab:kalabity_traj_sum}
	\end{table*}
	
	\begin{figure*}
		\centering
		\includegraphics[width=\linewidth]{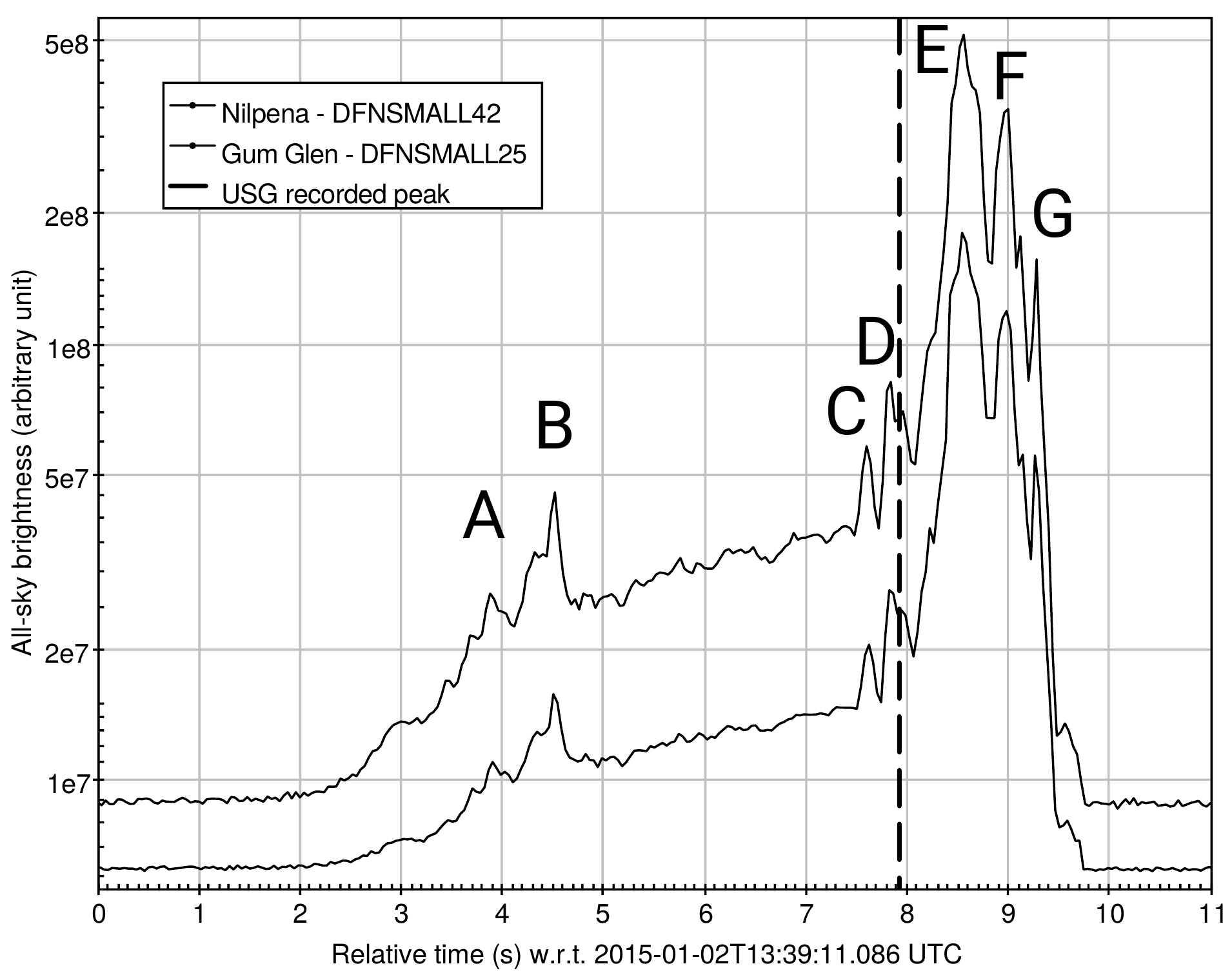}
		\caption{All-sky brightness (sum of all the pixels) from the Kalabity fireball, as recorded with the video cameras at the Gum Glen and Nilpena observatories. Using traditional PSF photometry on star Sirius the light curve is corrected to take into account the effect of auto-gain. The Nilpena curve has been shifted up for clarity. The peak brightness time recorded by the USG sensors (rounded to the nearest second) is marked by a vertical line.}
		\label{fig:kalabity_radiometric_light_curve}
	\end{figure*}

	The all-sky light curves display early fragmentation events under 0.05 and 0.08\,MPa and (peaks A and B in Fig. \ref{fig:kalabity_radiometric_light_curve}).
	The following part of the light curve is uneventful until the body encounters an order of magnitude higher dynamic pressures that eventually almost entirely destroys it (peaks C to G in Fig. \ref{fig:kalabity_radiometric_light_curve}).
	This adds to the list of large meteoroids \citep{2011M&PS...46.1525P} that undergo fragmentation under pressures several orders of magnitude smaller than the surviving material tensile strength on the ground, or pressures required to destroy the body in our case.
	
	We note that the time reported by the USG sensor (2015-01-02T13:39:19\,UTC) is in good agreement with the brightest peak (E) in our light curve determined to be 0.6\,s later (Fig. \ref{fig:kalabity_radiometric_light_curve} and Tab. \ref{tab:kalabity_traj_sum}).
	However the reported altitude is 38\,km. This does not correspond to our brightest peak E at 40.2\,km, but rather to the end of the very bright phase (peak G).

	Only 6 shutter breaks are resolvable on the image after the explosion on the Billa Kalina image, all $<4\degr$ on the horizon.
	Using the particle filter method of \citet{2017AJ....153...87S} on these data, we find that the main mass at this stage was only a couple of kilograms at the most.
	We are only able to track down to 33.4\,km at 8.4\,$\mbox{km s}^{-1}$. 	We suspect that this main mass is not visible down to ablation speed limit ($\simeq 3\mbox{km s}^{-1}$), because of a sensitivity issue: at this stage the meteoroid is at a large distance from the observatory ($>360$\,km), observed on an extreme elevation angle ($\simeq 3.5\degr$), and the sky background is unusually bright because of the light from the main explosions (peaks E-G) raising the background.
	We suspect the reason this feature is not visible on the closer Gum Glen image is because of the presence of clouds in the direction of the fireball, which efficiently scattered the light from the explosion and subsequently saturated the sensor on a much larger area than for Billa Kalina.

	The particle filter method of \citet{2017AJ....153...87S} can also be used to put a lower bound on the initial mass of the meteoroid.
	The near lack of deceleration before the main explosion implies that the mass to cross-section area ratio was large.
	Using reasonable assumptions on shape (spherical), and density ($\rho=3500\,\mbox{kg m}^{-3}$, chondiric), we find that the meteoroid was $>2600$\,kg ($> 1.1$\,m) before impact.
	We note that this assumes that the meteoroid is a single ablating body before the airbursts (peaks E-G).
	We know this assumption not to be well-founded because some fragmentation happened early on (peaks A and B in Fig. \ref{fig:kalabity_radiometric_light_curve}), explaining why this number is given as a lower limit.
	
	Using the velocity calculated at the brightest instant on DFN data (peak E in Tab. \ref{tab:kalabity_traj_sum}), and the impact energy measured by the USG sensors (Tab. \ref{table:usg_data}), we derive a 3400\,kg mass for this meteoroid, roughly equivalent to a 1.2\,m diameter body, larger than the \citet{2016Icar..266...96B} estimate because of a different impact speed used.
	
	The DFN dynamic initial size ($>1.1$\,m) is in good agreement with the USG photometric mass ($1.2$\,m).

\begin{figure*}
		\centering
		\includegraphics[height=0.9\textheight]{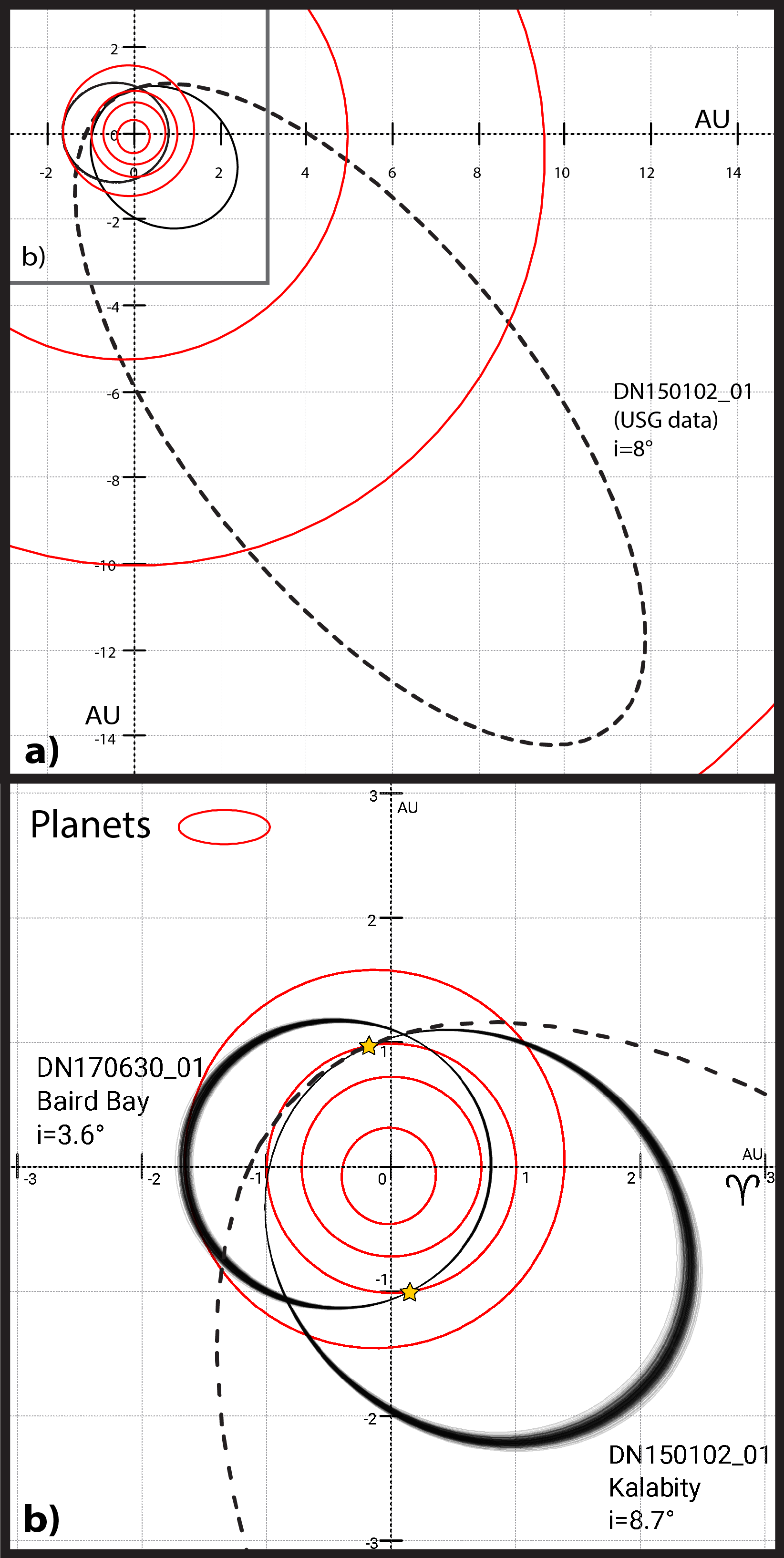}
		\caption{Ecliptic plot of the pre-atmospheric orbit of the Kalabity and Baird Bay meteoroids.
			\textbf{b} is limited to the inner solar system, while \textbf{a} goes out all the way to the orbit of Uranus.
			The solid lines are orbits using DFN data (the shades of grey in \textbf{b} represent the confidence region as calculated by Monte Carlo simulations), whereas the dashed lines are using USG data. The orbit of Baird Bay calculated from USG data is indistinguishable from the DFN one. On the other hand the orbit of Kalabity is very different, mostly because of a speed issue with USG data.}
		\label{fig:orbit}
	\end{figure*}

	The orbit of Kalabity is a typical main belt one with a semi-major axis of $1.80$\,AU (Tab. \ref{tab:orbit} and Fig. \ref{fig:orbit}), very different from the HTC type orbit derived from USG data (Tab. \ref{table:usg_data}).

	\begin{table*}
		\centering
		\begin{tabular}{cccc}
			\hline 
			parameter & unit & DN150102\_01 Kalabity & DN170630\_01 Baird Bay\\
			\hline 
			Epoch & TDB &   2015-01-02T13:39:11 &  2017-06-30T14:26:41 \\
			$a$ & AU & 1.80 $\pm$ 0.02 & 1.23 $\pm$ 0.01 \\
			$e$ & &  0.498 $\pm$ 0.006 & 0.35  $\pm$ 0.01 \\
			$i$ & \degr   & 8.73 $\pm$ 0.02 & 3.57  $\pm$ 0.05 \\
			$\omega$ & \degr  & 219.8 $\pm$ 0.09 & 259.06  $\pm$ 0.07 \\
			$\Omega$ & \degr  &  281.619 $\pm$ 0.001 &  98.801   $\pm$ 0.002 \\
			$q$ & AU &  0.908 $\pm$ 0.001 & 0.805  $\pm$ 0.004 \\
			$Q$ & AU & 2.70 $\pm$ 0.04 & 1.66 $\pm$ 0.03 \\
			$\alpha _g$ & \degr   & 64.3 $\pm$ 0.1  & 272.14  $\pm$ 0.02\\
			$\delta _g$ & \degr  & 51.7 $\pm$ 0.2 & -12.5 $\pm$ 0.1 \\
			$V _g$ & $\mbox{m s}^{-1}$ & 10776 $\pm$ 115 &  10007 $\pm$ 260 \\
			$T _J$ & & 3.89  & 5.14 \\
			\hline
			$\alpha _{inf}$ &  \degr    & 72.01 $\pm$ 0.03 & 273.29 $\pm$ 0.02 \\
			$\delta _{inf}$ & \degr  & 38.30 $\pm$ 0.02 & -15.88 $\pm$ 0.01 \\
			\hline 
		\end{tabular}
		\caption{Estimated orbital elements of DN150102\_01 Kalabity and  DN170630\_01 Baird Bay, with $1\sigma$ formal uncertainties. (equinox \textit{J2000}).}
		\label{tab:orbit}
	\end{table*}

	\subsection{DN170630\_01 - Baird bay}

	\begin{table*}
		\caption{Locations and nature of instrumental records DN170630\_01. P: Photographic record (long-exposure high resolution image), V: compressed PAL video (25 frames per second). $PW$ and $PF$ designate the de Bruijn encoding method, as described in Sec. \ref{sec:methods_dfn}. Ranges are from the fireball at 70\,km altitude. Photographic imaging system was out of order for O'Malley. Note that the Mulgathing camera did not receive the $PF$ firmware update immediately because of a temporary internet connectivity issue.}              
		\label{table:stations_BairdBay}      
		\centering                                      
		\begin{tabular}{c c c c c c}          
			\hline\hline                        
			Observatory & Instruments & Latitude & Longitude & Altitude (m) & Range (km) \\
			\hline      
			Mount Ive - DFNSMALL62 & P$_{PF}$ &  32.45919 S & 136.10332 E & 293  & 201  \\ 
			Days Hill - DFNEXT005 & P$_{PF}$ &  34.20749 S & 138.66151 E &  363 & 439  \\ 
			Nilpena - DFNSMALL12 & P$_{PF}$ & 31.02328  S & 138.23260 E & 122  & 447  \\
			Glenrest - DFNSMALL06 & P$_{PF}$ &  33.01963 S & 138.57554 E & 722  & 414  \\
			Billa Kalina - DFNSMALL43 & P$_{PF}$ &  30.23759 S & 136.51566 E & 113  & 387  \\
			Mulgating - DFNSMALL15 & P$_{PW}$ &  30.66078 S & 134.18608 E & 149	  & 274  \\
			Woomera - DFNSMALL14 & P$_{PF}$ &  31.19609 S & 136.82682 E & 163  & 329  \\
			O'Malley - DFNSMALL40 & V &  30.50663 S & 131.19534 E & 117  & 410  \\	
			\hline                                             
		\end{tabular}
	\end{table*}

	\begin{figure*}
		\centering
		\includegraphics[width=\linewidth]{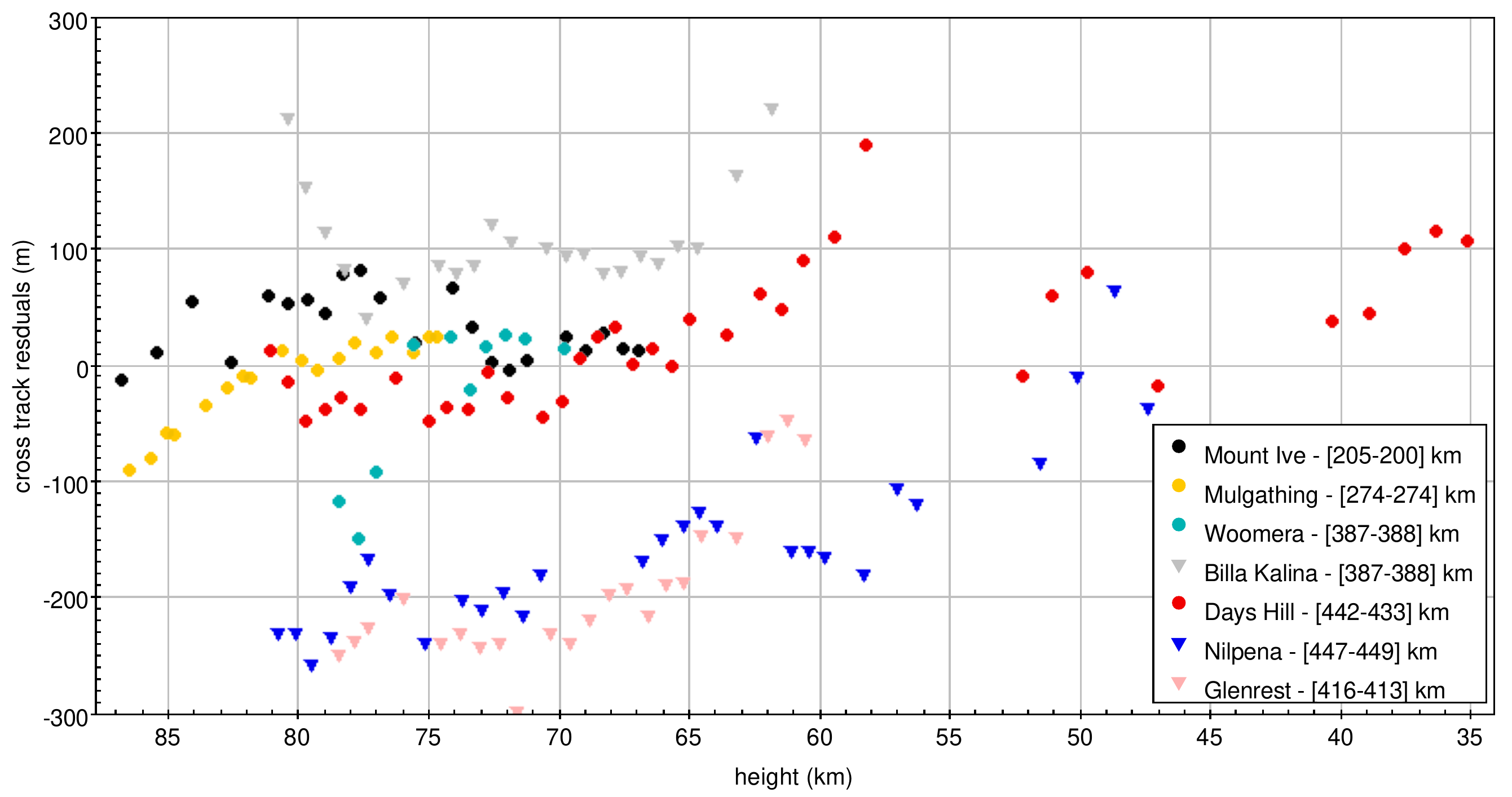}
		\caption{DN170630\_01 Baird bay. Cross-track residuals of the straight line least squares fit to the trajectory from each view point. These distances correspond to astrometric residuals projected on a perpendicular plane to the line of sight, positive when the line of sight falls above the trajectory solution. The distances in the legend correspond to the observation range [highest point - lowest point].}
		\label{fig:bairdbay_residuals}
	\end{figure*}
	
	\begin{table*}
		\centering
		\begin{tabular}{lcccccc}
			\hline
			Event & Time &  Speed &  Height &  Longitude &  Latitude &  Dynamic pressure \\
			& s & $\mbox{m s}^{-1}$ & m & \degr E & \degr N & MPa \\
			\hline
			Beginning & 0.0 & 15095$\pm$61 & 86782 & 134.23858 & -32.99306 & \\			
			A & 2.51 & 14906 & 52111 & 134.21168 & -33.08981 & 0.08 \\
			B & 3.51 & 13786 & 38817 & 134.20123 & -33.12718 & 0.42 \\
			C & 3.71 & 13140 & 36240 & 134.19919 & -33.13445 & 0.58 \\
			last astrometric datapoint & 3.80 & 12783 & 35181 & 134.19836 & -33.13743 & 0.65 \\
			D - max & 4.61 & 9568* & 25648* & 134.19083* & -33.16432* & 2.31* \\
			End & 5.46 & &&&& \\
			\hline
		\end{tabular}
		\caption{Summary table of bright flight events for DN170630\_01 Baird Bay. Fragmentation event letters are defined on the light curve (Fig. \protect\ref{fig:bairdbay_radiometric_light_curve}). Times are relative to 2017-06-30T14:26:41.50\,UTC. * marks figures that have been extrapolated. The end parameters have not been extrapolated as it is not possible to know what mass is left after the large explosion (peak D), and how this mass decelerated.}
		\label{tab:bairdbay_traj_sum}
	\end{table*}
	
	\begin{figure*}
		\centering
		\includegraphics[width=\linewidth]{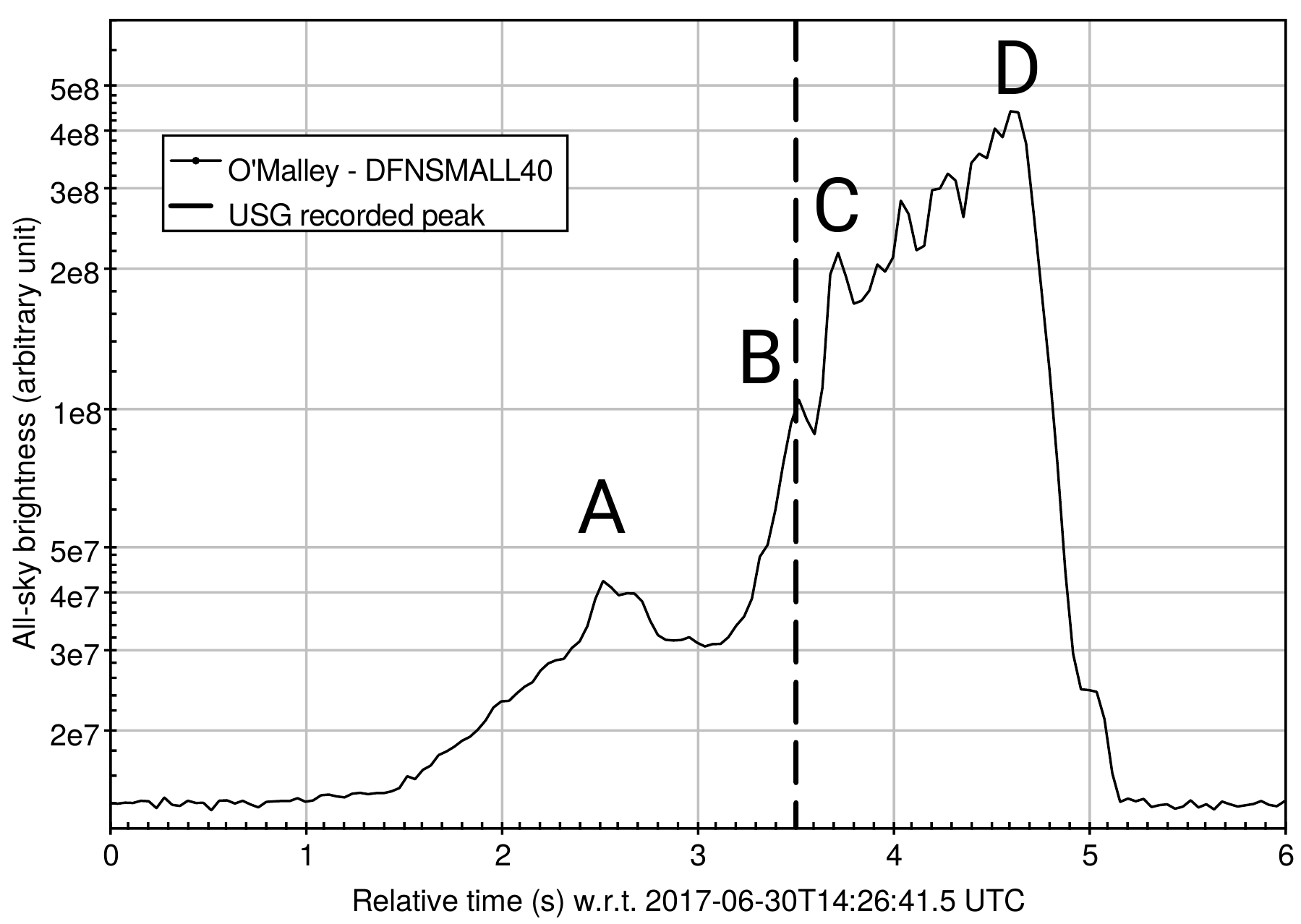}
		\caption{All-sky brightness (sum of all the pixels) from the Baird Bay fireball, as recorded with the video camera at the O'Malley observatory. Using traditional PSF photometry on star $\alpha$ Centauri the light curve is corrected to take into account the effect of auto-gain. The peak brightness time recorded by the USG sensors (rounded to the nearest second) is marked by a vertical line.}
		\label{fig:bairdbay_radiometric_light_curve}
	\end{figure*}
	
		The Baird Bay meteoroid entered the atmosphere on a very steep trajectory (72\degr to the horizon), on a trajectory that starts over land in Sceale Bay, and ended in the Southern Ocean $\sim$10\,km West of the Carca Peninsula (Fig. \ref{fig:images_map}).
	The bolide was visible from 2017-06-30T14:26:41.50\,UTC (3 minutes before midnight ACST) for 5.46\,s on the DFN camera systems (Tab. \ref{table:stations_BairdBay}).
	Several eye witnesses reported the bolide, notably from Adelaide, the closest densely populated area, 450\,km away.
	
	The closest DFN camera is Mount Ive station (190\,km away). The Mulgathing camera (250\,km directly North from the event) only caught the top of the fireball, as the image circle is cropped on the short side of the sensor (usually North and South).
	
	Like Kalabity, Baird Bay experienced early fragmentation under pressure $<1$\,MPa (peak A at 0.08\,MPa), however a much larger pressure was required to destroy it (peak D, most likely between 1 and 2\,MPa).
	
	Using the same particle technique as in Sec. \ref{sec:kalabity_results}, with a reasonable assumptions on shape (spherical), and density ($\rho=3500\,\mbox{kg m}^{-3}$, chondiric), we find that the meteoroid was $>9400$\,kg ($>1.7$\,m) before impact.	
	Using the particle filter we also find that the main mass was $\simeq7000$\,kg when it airburst at 26\,km altitude.
	Unfortunately no astrometric data is available after the airburst, as the only camera close enough to image the bolide at the end, Mount Ive, has a large area of the sensor saturated because of the airburst (peak D in Fig. \ref{fig:bairdbay_radiometric_light_curve}).
	The video record from the very distant O'Malley camera (410\,km) shows that some material was still ablating for at least 0.85\,s after the instant of peak brightness.
	This means that there is a distinct possibility that a main mass survived, and fell in the Southern Ocean, less than 10\,km from the coast off Point Labatt.
	
	The USG sensors locate the airburst $\lambda = 134.5 \degr$ $\phi = -34.3\degr$ (WGS84) at h=20\,km altitude (Tab. \ref{table:usg_data}).
	This position is $\sim$100\,km off to the South from our calculated entry parameters (Fig. \ref{fig:images_map}).
	
	On the other hand the USG geocentric velocity vector is consistent with our calculation. The radiant solutions are separated by only $0.4\degr$, and the speeds are different only by $0.1\,\mbox{km s}^{-1}$, in agreement within uncertainties.
	This implies that even with the wrong position, the orbit calculated from USG data (Tab. \ref{table:usg_data}) is in agreement with the DFN orbit (Tab. \ref{tab:orbit} and Fig. \ref{fig:orbit}).

	\section{Discussion}
	
	\subsection{Reliability of USG fireball data}
	
	We have compiled in Table \ref{table:usgscomp} how well USG events match independent observations of those events, using data both from the literature and the two fireballs described here.
	
	It is possible to discuss the reliability of the USG data in terms of different desired outcomes.

	\begin{table*}[h!]
		\caption{USG events that have their trajectory independently estimated. Note that the date/times of the events all match the independent measurements. The location is considered valid if the (latitude, longitude, height) is somewhere on the trajectory. 
        CSS: Catalina Sky Survey, V: video, P: photographic, PE: photometer, CV: casual video, I: infrasound. \ding{51}: correct within errors. $\thickapprox$: incorrect, but not far off. \ding{55}: incorrect. NR-U: not reported by USG. NR-L: not reported or yet published in literature.
        *: From light curve and infrasound data, [5] conclude that the impact kinetic energy for Ko{\v s}ice is $>0.1$\,kT, without an upper limit. References: 
        	(0) this work;
			(1) \citet{2018LPI....49.3006H}; 
			(2) \citet{2016Icar..266...96B};
			(3) \citet{2017P&SS..143..147B}; 
			(4) \citet{2013Natur.503..235B};
			(5) \citet{2013M&PS...48.1757B};
			(6) \citet{2017Icar..294..218F};
			(7) \citet{2010MsT.........27M};
			(8) \citet{2009Natur.458..485J};
			(9) \citet{2009A&A...507.1015B};
			(10) \citet{2018arXiv180105072P};
			(+) JPL Horizons ephemeris service, using CSS and ATLAS astrometry. 
        }              
		\label{table:usgscomp}      
		\centering                                      
		\begin{tabular}{c c c c c c c c c}          
			\hline\hline                        
			Event & Date (UTC) & Instruments & location & airburst height & speed & radiant  & energy & ref \\
			\hline    
			2018\,LA & 2018-06-02T16:44:12 & CSS & NR-L & NR-L & \ding{51} & \ding{51} & NR-L & + \\ 
			Crawford Bay & 2017-09-05T05:11:27 &   CV, I & $\thickapprox$ & \ding{51} & \ding{55} & \ding{55} & \ding{51} & 1 \\ 
			DN170630 - Baird Bay & 2017-06-30T14:26:45 & P, V & \ding{55} & \ding{55} & \ding{51}  & \ding{51} & \ding{51}  & 0 \\ 
			Dishchii'bikoh &  2016-06-02T10:56:32 & V, CV & \ding{55} & NR-U & NR-U &  NR-U & $\thickapprox$ & 10 \\ 	
			Romanian & 2015-01-07T01:05:59 & CV, PE, P &  \ding{51} &  $\thickapprox$  &  \ding{55} &  $\thickapprox$  & \ding{51} & 2, 3 \\ 
			DN150102 - Kalabity & 2015-01-02T13:39:10 & P, V & \ding{51} & \ding{51} & \ding{55} & \ding{55}  & \ding{51} & 0, 2 \\ 
			Chelyabinsk & 2013-02-15T03:20:21 & CV & $\thickapprox$ & \ding{55} & $\thickapprox$ & $\thickapprox$ & \ding{51} & 2, 4 \\ 
			Ko{\v s}ice & 2010-02-28T22:24:47 & V, P, PE & \ding{51} & $\thickapprox$ & \ding{51} & $\thickapprox$ & \ding{51}*  & 2, 5 \\ 
			Buzzard Coulee & 2008-11-21T00:26:40 & CV & \ding{51} & $\thickapprox$ & \ding{55}   & \ding{55} & NR-L  & 2, 7 \\ 
			Almahata Sita (2008 TC3)& 2008-10-07T02:45:40 & CSS & $\thickapprox$ & \ding{51} & \ding{55} & \ding{55} & \ding{51}  & 2, 6, 8, 9 \\ 
			
			\hline                                             
		\end{tabular}
	\end{table*}

	\subsubsection{For orbital studies}
	The factors that come into play to calculate a meteoroid orbit are the accuracy of the \emph{location}, the absolute \emph{time}, and the geocentric \emph{velocity vector}.
	
	All USG events in Table \ref{table:usgscomp} agree in absolute time with independent records to within a few seconds.
	
	Locations are correct in most cases, except for the Baird Bay event described in this work. However this $\sim$100\,km location issue is this case is not important for orbit calculation.
	
	Hence the questions lie with the 3 geocentric cartesian velocity components.
	\cite{2018Icar..311..271G} show that in most cases a precision of  $0.1\,\mbox{km s}^{-1}$ on the velocity is good enough for source region analysis, so we do not expect the lack of precision on the USG numbers  to be an issue here.
	An accurate height can be useful to take into account the deceleration in the atmosphere, but it is not essential as we are looking at massive bodies that hardly decelerate before the airburst.
	Because radiant and speed are less likely to be correlated than the cartesian velocity components, we have re-projected these velocity components as radiant and speed.
	The speeds are inconsistent in most cases (Tab. \ref{table:usgscomp}).
	The worst USG estimates are for the Buzzard Coulee meteorite (18.1$\,\mbox{km s}^{-1}$ calculated by \citet{2010MsT.........27M} compared to 12.9$\,\mbox{km s}^{-1}$ USG), 
	and the Romanian bolide (27.8$\,\mbox{km s}^{-1}$ calculated by \citet{2017P&SS..143..147B} compared to 35.7$\,\mbox{km s}^{-1}$ USG). These were underestimated by 28\%, and overestimated by 28\%, respectively.
	The USG radiant vector is off for most events, sometimes by only a couple of degrees (which does not drastically affect the orbit), but sometimes by as much as 90\degr (Buzzard Coulee and Crawford Bay events).
	From these considerations, only 4 out of 10 events in Table \ref{table:usgscomp} would have a reasonably accurate orbit if calculated from USG data: 2018\,LA, Baird Bay, Chelyabinsk, and Ko{\v s}ice.
	The USG orbits of some meteoroids are even misleadingly peculiar:
    Kalabity and Romanian would  be on unusual HTC orbits (as already noted by \citet{2016Icar..266...96B}).
	
	Therefore USG data can generally not be relied on for orbit determination, and there is no way to know for which events the data are reliable.

	\subsubsection{For material properties studies}
	The atmospheric behaviour of a meteoroid can yield some insights on what the meteoroid is made of and how it is held together.
	If no meteorite is recovered, the small set of USG sensors parameters contains very limited information regarding the rock itself, but it is nevertheless possible to derive the bulk strength of the body.
	An basic way of achieving this is to look at the dynamic pressure required to destroy the body (using $s = \rho_{atm} v^2$ from \citet{1981MoIzN....Q....B}).
	This is not a perfect indicator as it does not show subtleties in the rock structure, but it should be able to distinguish iron, chondritic, and cometary material, as these differ in bulk strengths by orders of magnitude.
	The key parameters are then the \emph{height} of peak brightness (to determine atmospheric density $\rho_{atm}$), and the \emph{speed} $v$.
	
	As shown by \citet{2016Icar..266...96B} (Tab. 4), the USG sensors tend to report reasonably accurate heights of peak brightness.
	We note that most of height inconsistencies are usually due to another peak in the light curve being recorded.
	
	As seen in the previous paragraph, speeds can be wrong by as much as 28\%, which induce a factor of 2 error in strength.
	We conclude that the inaccuracy of USG numbers do not affect strengths by more than an order of magnitude, this is good enough with respect to our original aim.

	\subsubsection{For size-frequency studies}
	The USG data have the advantage of using the entire planet as a collector, yielding large sample sizes that ground-based networks will never be able to reach for this class of objects.
	Hence they can be a good tool for size-frequency studies, provided the \emph{size} of the impacting bodies can be accurately determined, and the \emph{detection efficiency} is well constrained.
	
	As detailed in Sec. \ref{sec:usg_methods}, using the empirical relation of \citet{2002Natur.420..294B} and assuming a density, the radiated energy reported by the USG sensors can be converted into mass and size, with the caveat of speed accuracy.
	The energy estimates seem to match independent observation for the events presented here (Tab. \ref{table:usgscomp}).
	
	As of the detection efficiency, \citet{2002Natur.420..294B} mentions a 60-80\% Earth observation coverage by the USG sensors for their study on 1994-2002 data.
	If we subset the USG events in two different groups, before and after the study of \citet{2002Natur.420..294B}, we get on average 19 events per year before, and 26-27 events per year after September 2002.
	This 40\% increase would suggest a 100\% Earth coverage after 2002.
	However it is interesting to note that the $0.4$\, kT impact of \textit{2014 AA} \citep{2016Icar..274..327F} was not reported by the sensors.
	The distribution falls off near $0.1$\,kT TNT energy, so we only consider brighter $>0.1$\,kT events here.
	
	USG data is therefore useful for size frequency studies (like the work done by \citet{2002Natur.420..294B}, \citet{2013Natur.503..238B}, as long as the sub-population grouping is done by other means than by the orbit calculated using the USG velocity data.

	\subsubsection{For meteorite searching}
	Although metre-scale impactors are usually too big to be able to decelerate enough before reaching dynamic pressures that destroy them, these objects still have a large chance of surviving as meteorites. We try to assess here the viability of initiating dark flight calculations using a weather model combined the USG entry vector.
	All the parameters in Tab. \ref{table:usgscomp} (apart from time) need to be accurate.
	
	Although the height of peak brightness is wrong for Chelyabinsk, the reported \textit{(latitude, longitude, height)} triplet is located near the ground truth track, hence the fall analysis would not significantly change for large masses.
	Therefore of the events compiled in Table \ref{table:usgscomp}, only 2 out of 9 events (Ko{\v s}ice and Chelyabinsk) would have reasonably accurate fall positions if computed from USG records.
	
	But even worse, the 0.1\degr error on latitude/longitude translates into a $\pm$5\,km error on position on the ground, this is particularly large for undertaking meteorite searching activities.
	
	From these considerations, it would be ill-advised to undertake meteorite searching solely based on USG data.

	\subsection{On the ground-based imaging capabilities of metre-scale impactors}
	With the help from collaborators outside Australia, the DFN is expanding into the Global Fireball Observatory, and will eventually cover 2\% of the Earth surface in the next few years.
	Metre-scale object will fall on the covered area every 1-2 years on average, but is the currently deployed technology fit to observe such events?

	\subsubsection{Night time observations}
	Fireball observatories are typically optimised to observe the behaviour of macroscopic meteorite droppers throughout their trajectory during the night.
	The challenge is mostly a dynamic range one: being sensitive enough to observe the smaller meteoroid at a high altitudes to get precise entry speed for orbit calculation, whilst not saturating the records of larger rocks shining 100 million times brighter when they reach the dense layers of the atmosphere.

	So far no iron meteorite fall has been instrumentally observed, but it is expected that this class of objects contains the smallest meteoroids (ie. the faintest fireball) that can drop a meteorite, as their large strength allows them to enter with limited mass loss due to fragmentation.
	For instance, if we assume little to no gross fragmentation \citep{1994A&A...292..330R}, to produce a 100\,g meteorite the parent meteoroid ($\rho=7900\,\mbox{kg m}^{-3}$) can be as small as 0.5\,kg $\equiv$ 5\,cm diameter, assuming the most favourable entry conditions (vertical entry at $12\,\mbox{km s}^{-1}$).
	It is desirable to observe the meteor before the rock starts being affected by the atmosphere too much, 80\,km, altitude at which it would glow at magnitude $M_V$=-1.5 (assuming a luminous efficiency of 0.05).
	
	On the bright end, we look at the compilation of \citet{2015aste.book..257B} and see that metre-scale events usually approach $M_V^{max} = -18$, although this is highly dependent on their atmospheric behaviour, where and how important the fragmentation events are.
	
	The set goal is then to have instruments that can cover 20 stellar magnitudes of effective dynamic range.

	Long exposure high resolution fireball camera systems have a long track record for yielding meteorite ground locations and orbits (listed as "dedicated search from detailed computation of trajectory" by \citet{2015aste.book..257B}), compared to video systems.
	Thanks to their logarithmic response, film based imagers cover a very wide dynamic range ($\sim$15 stellar magnitudes), but those systems are costly and impractical for large distributed autonomous fireball networks \citep{2017ExA...tmp...19H}, and do not achieve the 0 magnitude sensitivity objective.
	The DFN \citep{2017ExA...tmp...19H} and the European Network \citep{2016LPICo1921.6221S} have recently switched from film to digital camera technology.
	This shift has simplified some operational aspects (eg. enhanced autonomy, better reliability, eased data reduction), but it has come at the cost of a much limited dynamic range: $\sim$8 magnitudes without saturation. For astrometric purposes this range can be extended to 15 magnitudes \citep{2018arXiv180302557D}, but this is still quite far from the 20 magnitudes objective.
	
	Video cameras are generally more sensitive than the still imagers, but suffer from the same limited dynamic range.
	Although a lot of events have been recorded, fixed frame rate TV systems have not been proficient in yielding meteorite fall positions.
	This is likely to be due to the low resolution offered by those systems (a PAL video system with a matching circular fisheye lens has an average pixel size over $10\times$ larger than the DFN cameras'), and the difficulty of getting enough stars for astrometric calibration across the field of view (most of these cameras cannot shoot long exposures).
	However recent advances in digital video camera technology allow higher resolutions, long exposures for calibration, and higher bit depth, so we expect networks based on these systems to be more successful at meteorite recovery in the near future (eg. the Fireball Recovery and InterPlanetary Observation Network (\textit{FRIPON}) network of \citet{2015pimo.conf...37C}).

	\subsubsection{Day time observations}
	
	The easy exposure control on industrial digital cameras allows low-noise long exposure calibration shot to be taken at night, but also permits very short exposures to operate during the day.
	The \textit{FRIPON} network endeavours to operate their cameras during both nighttime and daytime \citep{2014pim4.conf...39A}, however fireball detection on daytime frames appears somewhat challenging \citep{2016pimo.conf...73E}.
	Even if calculating fall positions turns out to be difficult from daytime data, the prospects of being able to calculate orbits for meteorites that have been independently recovered are very interesting (9 out of 14 US meteorite falls in the last 10 years do not have a trajectory solution published), as the astrometric calibration of casual footage can be very time consuming.

	\section{Conclusions}
	
	This work investigates the Near-Earth Objects (NEO) impacting population around the metre-scale size range.
	Such events are relatively rare (35-40 per year), therefore a large collecting area is crucial in order to study them. 
	The Desert Fireball Network (DFN) is leading the effort as a ground based instrument, covering over 3 million km$^2$.

	Meteoroids that have been observed by both the USG sensors and independent means comprises a small set of 9 events. 
	In this study we use a precise comparison of these events to assess the reliability of the USG sensors for NEO studies, yielding the following unequivocal conclusions:

	\begin{enumerate}
		\item USG sensors data are generally unreliable for orbit calculations. The new metre-scale impactors source region of \citet{2016Icar..266...96B} (Halley-type comet orbits) is based on 3 particular USG meteoroid orbits.
		We have shown that 2 of these are erroneous, seriously questioning the existence of this source region.
		
		\item Size frequency distribution work relies on determining rough sizes and having a good knowledge of the probing time-area. The USG seem to achieve both with reasonably good precision.
		This confirms the sound basis of the work done by \citet{2002Natur.420..294B} and \citet{2013Natur.503..238B}.
		
		\item Basic impactor physical properties (size and strength) can be well constrained with USG data.
		This validates the conclusions of \citet{2016Icar..266...96B} that relate to physical properties of objects. 
		
		\item Based on how often the derived trajectories are wrong, it would be naive to invest large amounts of resources to undertake meteorite searching using USG data.
	\end{enumerate}
	
	We also note that ground based fireball networks must find solutions to increase the dynamic range of their observations, in order to get sound observation data when metre-scale objects impact the atmosphere.

\section*{Acknowledgements}

This research is supported by the Australian Research Council through the Australian Laureate Fellowships and Discovery Proposal schemes, receives institutional support from Curtin University, and uses the computing facilities of the Pawsey supercomputing centre.
		The DFN data reduction pipeline makes intensive use of Astropy, a community-developed core Python package for Astronomy \citep{2013A&A...558A..33A}.

\bibliographystyle{mnras}
\bibliography{research} 

\bsp	
\label{lastpage}
\end{document}